\documentclass{aastex}
\usepackage[twocolumn]{emulateapj5}

\usepackage{epsfig}
\usepackage{apjfonts}

\submitted{ApJL, accepted}

\def\ltsima{$\; \buildrel < \over \sim \;$}
\def\simlt{\lower.5ex\hbox{\ltsima}} 
\def\gtsima{$\; \buildrel > \over \sim \;$}
\def\simgt{\lower.5ex\hbox{\gtsima}} 

\shorttitle{Radio Jet in z=4.3 Quasar} 
\shortauthors{Cheung}

\begin{document}

\title{Radio Identification of the X-ray Jet in the z=4.3 Quasar 
GB~1508+5714}

\author{C.~C. Cheung}
\affil{Department of Physics, MS~057, Brandeis University, Waltham, MA
02454}
\email{ccheung@brandeis.edu}

\begin{abstract}

The recent discovery of an X-ray jet in the z=4.3 quasar GB~1508+5714 by
Yuan et al. (astro-ph/0309318) and Siemiginowska et al. (astro-ph/0310241)
prompted a search for its radio counterpart. Here, we report the
successful discovery of faint radio emission from the jet at 1.4 GHz using
archival VLA data. The X-ray emission is best interpreted as inverse
Compton (IC) emission off the CMB as discussed by the previous
investigators. In this scenario, its high X-ray to radio monochromatic
luminosity ratio, compared to previously detected IC/CMB X-ray jets at
lower redshift, is a natural consequence of its high redshift.

\end{abstract}


\section{Background}

Since its launch in 1999, the Chandra X-ray Observatory has been used to
detect a large number of X-ray jets in Active Galactic Nuclei (AGN), where
prominent radio jets were previously known to exist \citep[see e.g.,][and
associated website\footnote{http://hea-www.harvard.edu/XJET/}]{har02a}.  
The recent report \citep{yua03,sie03b} of an extended X-ray jet
originating from the z=4.3 quasar GB~1508+5714, where previous
observations showed no obvious sign of extended radio emission, presents
an interesting case. The X-ray feature is strong -- well over 100 counts
were detected from it in the $\sim$90 ksec Chandra exposure. An archival
HST image helps rule out the possibility that it is due to a foreground
galaxy or a gravitationally lensed image of the quasar \citep{sie03b}.
Based on deep X-ray source counts, it has a low probability of being a
random unassociated X-ray field source.

As discussed by the previous authors, such detections of X-ray jets at
large redshifts are actually to be expected as a natural consequence of
the inverse Compton (IC) off the CMB model \citep[e.g.,][]{tav00,cel01}.  
This is because the (1+z)$^{4}$ dependence of the CMB energy density
compensates for cosmological dimming of radiation, so that IC/CMB X-ray
jets should remain detectable out to large cosmological distances
\citep{sch02a}. The model has been successfully applied to account for
X-ray jets in many other powerful quasars at more modest redshifts
\citep[e.g.,][]{sam02}, requiring that the jets are still highly
relativistic on kilo-parsec scales, in order that the electrons in the jet
frame see an adequately boosted photon source. However, the lack of a
detection of the GB~1508+5714 jet at lower frequencies, along with only a
rough constraint on the X-ray spectrum, could not rule out a synchrotron
origin for the X-rays \citep{sie03b}. A previous search in the radio for a
proposed X-ray jet in another high redshift (z=5.99) quasar,
SDSS~1306+0356 \citep{sch02b}, yielded only an upper limit of $<$0.1 mJy
at 1.4 GHz \citep{pet03,sch03}.

Distinguishing between the two possible emission processes is important,
as they probe different energetic phenomena. In the case of synchrotron
X-ray emission, the X-rays mark sites of particle acceleration with
electrons accelerated up to $\gamma\sim10^{7}$ with very short lifetimes
\citep[e.g. M87;][]{har03}. When considered together with minimum
energy/equipartition conditions, the IC/CMB model offers important
constraints on the beaming, magnetic field, and jet power
\citep{tav00,cel01}.

Both the IC and synchrotron interpretations of X-ray jet emission usually
require a population of relativistic electrons emitting synchrotron
radiation at lower (radio) frequencies.  Based on the X-ray flux and
spectrum, the two models give different predictions of the radio component
flux and spectrum. Here, we report the detection of such a radio component
coincident with the X-ray feature extending from GB~1508+5714 from an
analysis of archival VLA data. This supports its interpretation as an
X-ray jet and we discuss the X-ray emission in the context of the new 1.4
GHz detection, along with the previously set optical, and a new 8.4 GHz
limit. Following \citet{yua03} and \citet{sie03b},
H$_{0}=71~$km~s$^{-1}$~Mpc$^{-1}$, $\Omega_{\rm M}=0.27$ and $\Omega_{\rm
vac}=0.73$ \citep{spe03} are assumed throughout, so 1\arcsec\ = 6.871 kpc.

\section{Archival Radio Observations}

Two datasets which utilized the VLA in its highest resolution
A-configuration were obtained from the NRAO\footnote{The National Radio
Astronomy Observatory is a facility of the National Science Foundation
operated under cooperative agreement by Associated Universities, Inc.}
archive. The data were calibrated in the NRAO AIPS package \citep{bri94a}
and brought into DIFMAP \citep{she94} for editing, self-calibration, and
imaging. The flux density scale was set on the VLA 1999.2 scale using
scans of 3C~286 as outlined in the VLA Calibrator Manual \citep{per03}.

\begin{figure*}
\figurenum{1}
\begin{center}
\epsfig{file=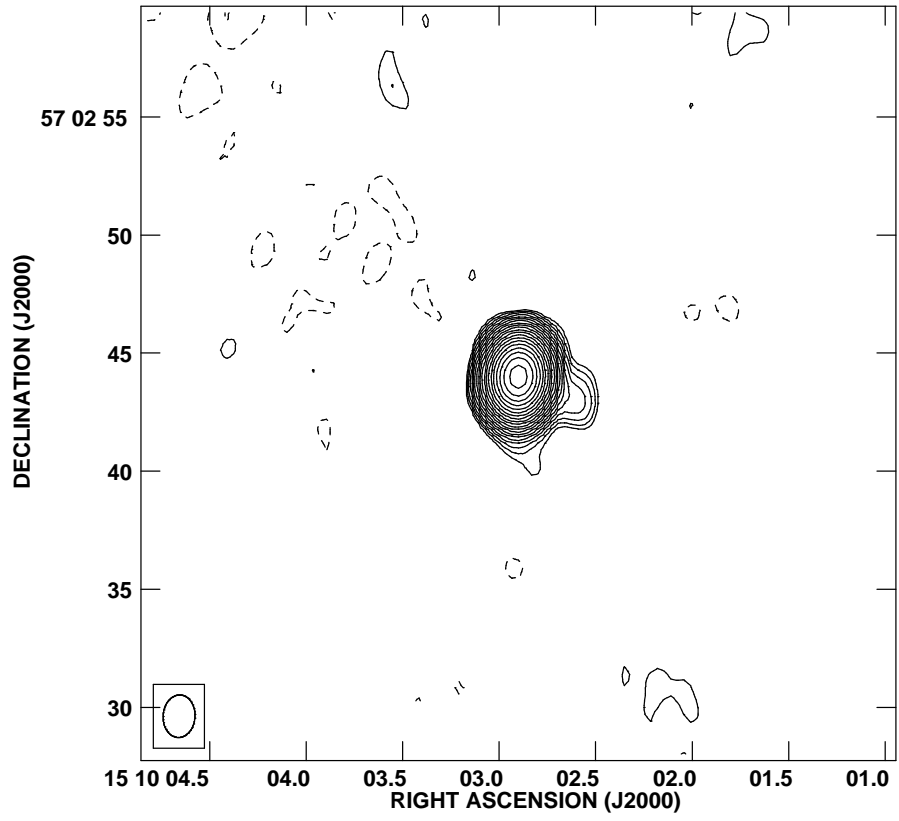,width=5.5in,angle=0}
\end{center}
\figcaption[f1.eps]{\label{fig-1} Naturally weighted VLA 1.4 GHz archival
snapshot image of GB~1508+5714 (at center) and its $\sim$2.5$\arcsec$ long
radio jet extending south of west of the quasar. The contours begin at
0.25 mJy/beam (2 times the measured RMS in the image) and the positive
values (solid contours) are spaced by factors of $\sqrt{2}$, with an image
peak of 224 mJy/beam. The restoring beam is plotted at the bottom left
corner and has dimensions 1.8\arcsec$\times$1.35\arcsec\ at a position
angle of --3.61$^{\circ}$.}
\end{figure*}

The 1.4 GHz dataset -- a single five minute snapshot observation --
yielded a 10$\sigma$ detection of a 1.2 mJy feature extended from the
quasar (Figure~\ref{fig-1}).  This feature is only about $0.5\%$ of the
core flux ($224\pm5\%$ mJy), so it is no surprise that the original
investigators \citep{mor97} simply concluded that the quasar was an
unresolved point source from these data. The measured off-source RMS of
$\simlt$0.12 mJy in the image is within 50$\%$ of the thermal noise limit
expected from the integration time. The resultant image has a dynamic
range of almost 2,000:1.

\begin{figure*}
\figurenum{2}
\begin{center}
\epsfig{file=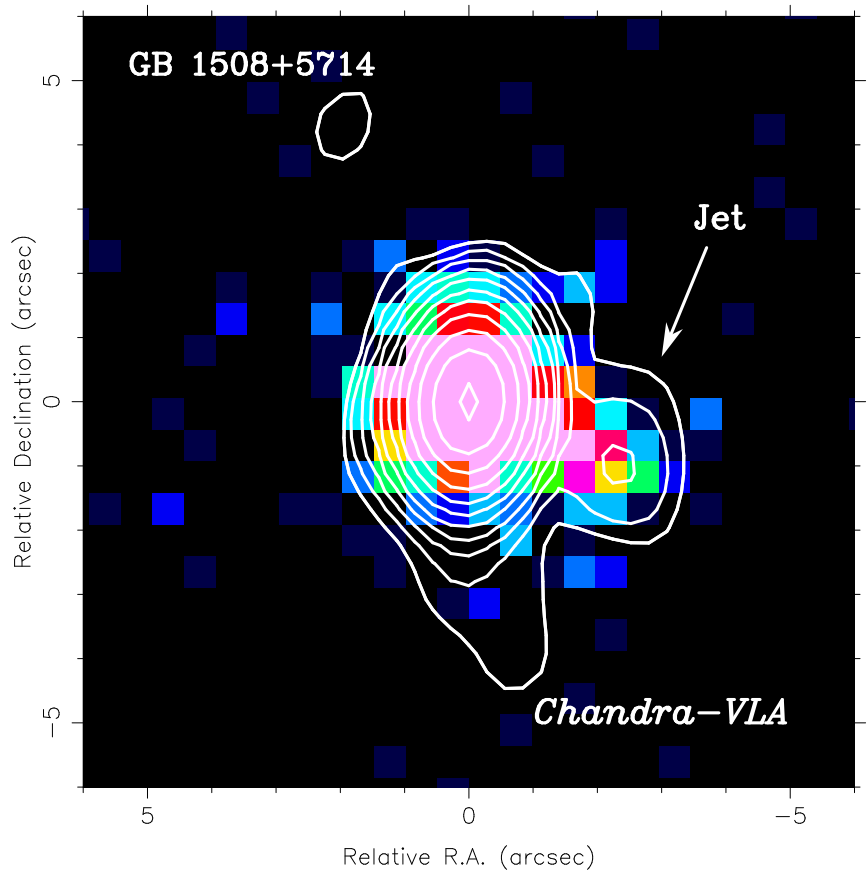,width=5.5in,angle=0}
\end{center}
\figcaption[f2.eps]{\label{fig-2} Chandra X-ray image of GB~1508+5714
(color) from the archive data published in \citet{yua03} and
\citet{sie03b}, with VLA 1.4 GHz image overlaid. The quasar is at the
origin, and the jet feature indicated. The radio image is from the same
data presented in Figure~\ref{fig-1} but restored with the uniformly
weighted beam (1.52\arcsec$\times$1.03\arcsec\ at PA=--4.44$^{\circ}$).
The lowest contour plotted is 0.2 mJy/beam and subsequently spaced by
factors of two.}
\end{figure*}

The jet knot was modeled with an elliptical gaussian profile in both the
(u,v)  plane with DIFMAP's \verb+MODELFIT+ utility, and in the map plane
with the JMFIT task in AIPS. The fits gave consistent measures of the
deconvolved size: $1\arcsec\times<1\arcsec$, elongated along the jet
direction.  We note that this is comparable to the beamsize for this
observation so the source is consistent with being unresolved, and the
size should be considered an upper limit. The separation from the core is
2.5\arcsec\ at a position angle of $-114^{\circ}$, and is comparable to
that measured from the X-ray data. Interestingly however, the peak in the
X-ray jet appears somewhat closer to the nucleus than the radio centroid
in the image overlay (Figure~\ref{fig-2}), by about one Chandra ACIS pixel
(0.492\arcsec\ per pixel), or equivalently $\sim$3 kpc projected distance.
The images were aligned by the cores to better than half of a pixel.
\citet{sie03b} found the peak in the X-ray jet to be $\sim$2\arcsec\ away
(see also their subpixel re-binned radial profile published in their
Figure~2), consistent with our qualitative assessment. \citet{yua03}
stated the same peak to be $\sim$3\arcsec\ distant from the nucleus,
although we measured a smaller value off their Figure~1. Much more
apparent X-ray/radio offsets are seen in other powerful quasars
\citep[e.g. up to $\sim$2\arcsec\ in PKS~1127--145;][]{sie02}. Most, if
not all of the X-ray counts from the GB~1508+5714 jet lie within the
outermost radio contour in the VLA image (Figure~\ref{fig-2}).

A 10-minute 8.4 GHz observation obtained on Nov 3, 1996 (Program AD388)
yielded no detection of the radio jet. The measured off-source RMS in the
naturally weighted image ($\sim$0.35\arcsec\ beam) is at about the
expected thermal noise limit of 0.045 mJy. In order to judge if the
non-detection was a result of the greater resolution in this image
compared to the 1.4 GHz map, the 8.4 GHz dataset were tapered by different
amounts. No outstanding feature appeared above the residual artifacts from
the dirty beam. The measured RMS at the expected position of the jet in a
tapered image restored with a 0.75\arcsec\ beam is 0.3 mJy
(3$\sigma$). This, along with the 1.4 GHz detection, is consistent with an
$\alpha\simgt$0.8 radio spectrum ($F_{\nu}\propto\nu^{-\alpha}$), which
agrees with the measured X-ray spectral index: 0.9$\pm$0.36 \citep{sie03b}
and 0.92$_{-0.33}^{+0.38}$ \citep{yua03}.

\section{Constraints from the Radio Data}

\begin{figure*}
\figurenum{3}
\begin{center}
\epsfig{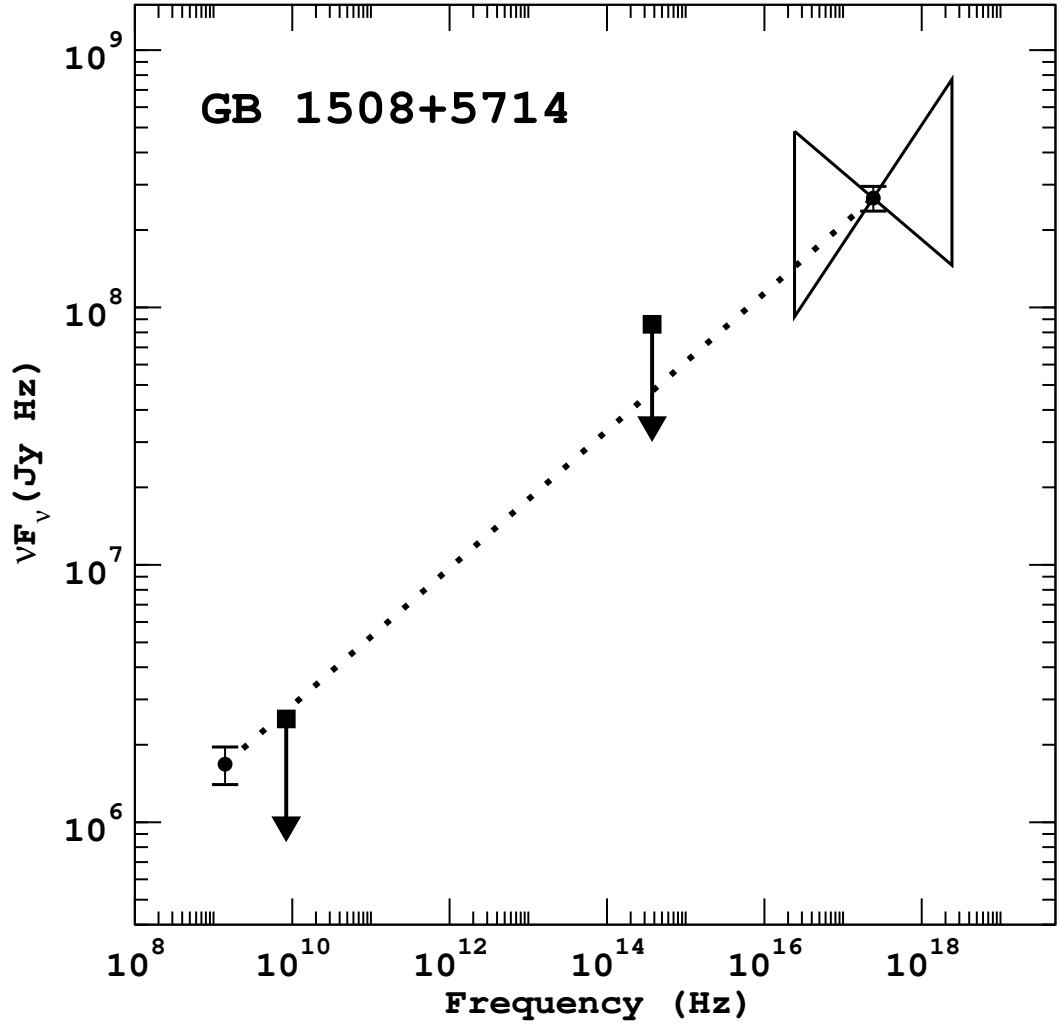}
\end{center}
\figcaption[f3.eps]{\label{fig-3} Radio-to-X-ray spectral energy
distribution of the GB~1508+5714 jet.  The bowtie indicates the X-ray
spectrum measurement with 1$\sigma$ uncertainty from \citet{sie03b} and
the arrows indicate 3$\sigma$ limits. The radio data are from this work,
and the X-ray point and optical limit were taken from \citet{sie03b}. A
dotted line shows the $\alpha_{\rm rx}$ slope and simply joins the radio
and X-ray detections.}
\end{figure*}

The spectral energy distribution of the jet knot in GB~1508+5714 is shown
in Figure~\ref{fig-3}. If the X-ray and radio emission are drawn from the
same population of relativistic electrons emitting synchrotron radiation,
the spectral index will be about what is measured between the 1.4 GHz and
X-ray (using 1.68$\times$10$^{-6}$ ph/cm$^{2}$/s/keV at 1 keV, reported by
\citet{sie03b}) detections: $\alpha_{\rm rx}$ = 0.73. This is a typical
value for the spectrum seen in radio jets \citep{bri84}, and is consistent
with the measured X-ray spectrum of $\simeq0.9\pm0.36$
\citep{yua03,sie03b}, and our constraint on the radio spectrum of
$\simgt$0.8 (the modest 8.4 GHz dataset does not preclude that the higher
frequency radio emission, $>$40 GHz in the source frame, was actually
resolved out). The optical HST limit (3$\sigma$) from \citet{sie03b} is
not useful in this case, as it hovers over the radio-to-X-ray spectrum,
and only weaker constraints are available at other optical bands
\citep{yua03}.

We can estimate an equipartition magnetic field of $B_{\rm
eq}\simeq9.7\times10^{-5}~\delta^{-1}~\gamma_{min}^{-0.12}~(\eta~/~f)^{0.27}$~G,
by adopting the observed $\alpha_{\rm rx}$ value as the optically thin
spectral index, and that the particles and field fill a fraction $f$ of a
2.9$\times$10$^{67}$ cm$^{3}$ sphere, corresponding to the size upper
limit derived from the radio detection.  The factor $\eta$ is the ratio of
magnetic field and electron energy densities, and $\gamma_{min}$ is the
lower energy cutoff of a power law distribution of relativistic particles.
If we further assume $\eta$=1, $f$=1, and $\gamma_{min}$=10, then $B_{\rm
eq}\simeq7.4\times10^{-5}~\delta^{-1}$ G, where $\delta$ is the unknown
jet Doppler beaming factor.  In this field, electrons emitting the highest
energy X-rays detected \citep[$\sim$5 keV;][]{sie03b} will have
$\gamma\sim10^{8}$ and short lifetimes ($\sim$50 years) -- this would
require continuous in-situ reacceleration of high energy particles in
order that the X-ray jet not be a transient feature. At this high
redshift, the energy density of the CMB already exceeds that of the
equipartition field, even without bulk motion (see below), so inverse
Compton losses will already be dominant.  The lifetime of the
$\gamma\sim10^{8}$ electrons calculated above is then a strict upper
limit. Taking this evidence together, it is unlikely that the jet X-ray
emission is dominated by synchrotron losses.

An IC/CMB origin for this high redshift X-ray jet is preferred
\citep{yua03,sie03b}, and give us an additional constraint on the allowed
range of magnetic field and jet Doppler factor to the equipartition
condition. The following expression is taken from \citet{tav02}:

\begin{equation}\label{eqn-1}
B=\delta~[\frac{3}{2}
\frac{c(\alpha)}{\sigma_{T}~c}
\frac{1}{U_{rad}~\nu_{0}^{\alpha-1}}
\frac{\nu_{c}^{\alpha}}{\nu_{s}^{\alpha}}
\frac{F_{c}(\nu_{c})}{F_{s}(\nu_{s})}]^{-1/(1+\alpha)}, 
\end{equation}

\noindent where, $c(\alpha)$ is a dimensionless function of $\alpha$
\citep[e.g.,][]{ghi85}, $\sigma_{T}$ is the Thomson cross section, and
$U_{rad} = 4.19\times10^{-13} (1+z)^{4}$ ergs cm$^{-3}$ for the CMB
\citep[e.g.,][]{sch02a}, whose spectrum is peaked near
$\nu_{0}=1.6\times10^{11}$ (1+z) Hz \citep{har02a}. Assuming $\alpha=1$,
the observed ratio of the Compton (X-ray)  and synchrotron (radio) fluxes
($F_{c}$ and $F_{s}$, observed at $\nu_{c}$ and $\nu_{s}$, respectively),
allow us to estimate $B\simeq7.2\times10^{-6}~\delta$ G. The corresponding
equipartition calculation gives $B_{\rm eq}\simeq10^{-4}~\delta^{-1}$ G,
making the assumptions as above about $\eta$, $f$, and $\gamma_{min}$.
Both estimates of $B$-field are reasonably insensitive to the assumed
spectral index: up to $\sim$25$\%$ smaller for $\alpha=0.9$, the measured
X-ray spectral index. In order to reconcile the differences between the
two equations (opposite dependencies of $B$ on $\delta$), a Doppler factor
greater than one is required. These two constraints bracket a plausible
range of solutions around a 30$\mu$G field and $\delta\sim$4.  
\citet{sie03b} arrived at similar conditions using slightly different
assumptions. The unresolved nucleus is a variable X-ray and (flat
spectrum) radio source with a very high X-ray to optical luminosity
\citep[see,][and references therein]{mor97}. These observations have been
taken as evidence of beaming on smaller scales, in the nucleus.

\begin{figure*}
\figurenum{4}
\begin{center}
\epsfig{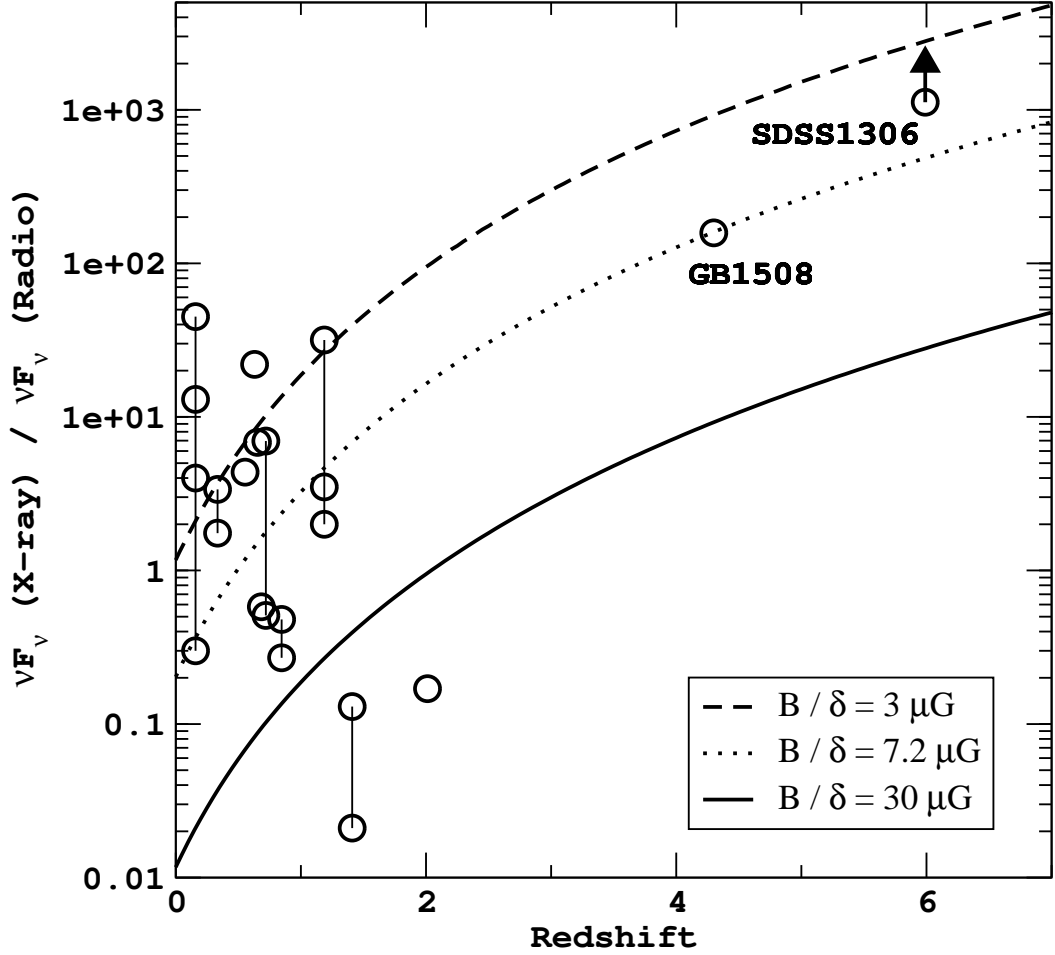}
\end{center}
\figcaption[f4.eps]{\label{fig-4} Plot of the ratio of the jet X-ray to
radio monochromatic luminosity vs. redshift. Only jet features interpreted
by the authors as IC/CMB X-ray emission are plotted. The curves indicate
the expected ratio for given combinations of $B$ and $\delta$, which scale
as (1+z)$^{4}$. For reference, the $B\sim~30\mu$G and $\delta\sim$4 case
derived for GB~1508+5714, which used the additional equipartition
constraint, defines the dotted line which lies in between the other two
curves. Light vertical lines connect features from the same source. The
data are for (in order of increasing redshift) 3C~273-A to D
\citep{sam01}, 1150+497-A, B, PKS~1136--135-B \citep{sam02}, B2~0738+313-A
\citep{sie03a}, PKS~0637--752-WK7.8 \citep{cha00}, 3C~207-knot
\citep{bru02}, PKS~1354+195-A, B, 3C~179-A, B \citep{sam02},
PKS~1127--145-A to C \citep{sie02}, Q0957+561-B, C \citep{cha02}, 3C~9-jet
\citep{fab03} with the corresponding radio flux (sum of knots F to J)
taken from \citet{bri94b}, and GB~1508+5714 and SDSS~1306+0356 (see text).  
}
\end{figure*}

The ratio of the monochromatic X-ray to radio luminosities (i.e., in
$\nu$F$_{\nu}$) of the GB~1508+5714 jet is 158. This is one of the highest
amongst the known X-ray jets which can be attributed to IC/CMB emission so
far. In Figure~\ref{fig-4}, this ratio is plotted vs. redshift along with
IC/CMB X-ray jets taken from the literature. The values seem to vary
widely over 4 orders of magnitude, but predominantly show X-ray to radio
ratios greater than 1. The jet in GB~1508+5714 sees 1--2 orders of
magnitude times greater energy density from the CMB than other jets at
lower redshift, so it is tempting to speculate that its extreme redshift
may account, to first order, for its large X-ray to radio luminosity
ratio. This could similarly, account for the large X-ray to radio ratio
limit obtained for the proposed X-ray jet in the z=5.99 quasar
SDSS~1306+0356 \citep[][Fig.~\ref{fig-4}]{sch02b,sch03}.

The observed monochromatic X-ray to radio luminosity ratios can be
compared to what is expected in the IC/CMB model. We can obtain an
expression for the expected ratio by rearranging equation~\ref{eqn-1} (and
setting $\alpha=1$):

\begin{equation}\label{eqn-2}
\frac{\nu_{c}}{\nu_{s}}
\frac{F_{c}(\nu_{c})}{F_{s}(\nu_{s})} =
\frac{4.19\times10^{-13} (1+z)^{4}}{(B/\delta)^{2}/8\pi}.
\end{equation}

\noindent The flux ratio is simply proportional to the ratio of the energy
densities in the CMB and magnetic field, with a Doppler factor term. This
expression is plotted in Figure~\ref{fig-4} for several different
combinations of $B$ and $\delta$, one of which, best suits the
observations of GB~1508+5714. It appears that varying $B$ and $\delta$ can
account for the large spread of observed X-ray to radio luminosity ratios
within a given redshift range. Larger $B$-field, or more likely, smaller
$\delta$ may account for the low values observed in Q0957+561
\citep[z=1.41,][]{cha02} and 3C~9 \citep[z=2.012,][]{fab03}, as they are
known to have low radio core dominance, and weak VLBI structures
\citep{cam95,hou02} compared to the other sources studied here. In all but
3C~179 \citep[z=0.846,][]{sam02} where multiple knot regions along the jet
can be distinguished, the X-ray to radio flux ratio decreases with
increasing distance from the nucleus, and can similarly be accounted for
by varying $B/\delta$. As discussed specifically in the case of 3C~273
\citep[z=0.158,][]{sam01}, the variations along the jet may indicate
deceleration, or increasing $B$ along the jet, possibly by compression in
strong shocks. Future detections in the redshift range $\sim$2--4 or
higher with large X-ray/radio flux ratios can lend further support to the
currently preferred IC/CMB model used to explain the X-ray emission in
quasar jets.

\acknowledgements

An anonymous referee is thanked for making the point that Compton losses
are already dominant in the synchrotron X-ray case, and several other
useful comments. The author is grateful to John Wardle for his advice and
encouragement throughout the course of this work, and to Dan Harris, Aneta
Siemiginowska, Dave Roberts, and Fabrizio Tavecchio for useful
discussions. Radio astronomy at Brandeis University is supported by the
National Science Foundation through grants AST 98-02708 and AST 00-98608.  
Additional support by NASA grant GO2-3195C from the Smithsonian
Astrophysical Observatory, and HST-GO-09122.08-A from the Space Telescope
Science Institute, which is operated by AURA, Inc., under NASA contract
NAS 5-26555 is acknowledged.


\end{document}